\documentclass[12pt,linenumbers, preprint]{aastex}

\usepackage{amsmath,longtable,natbib}
\usepackage{float}
\usepackage{hyperref}
\usepackage{adjustbox}
\usepackage{caption}
\usepackage{graphicx,epstopdf}
\DeclareGraphicsExtensions{.ps}
\shorttitle{Wide-band timing of GMRT MSPs}
\shortauthors{Sharma et al.}
\usepackage{subfig}
\begin{document}
\captionsetup[figure]{labelfont={bf},labelformat={simple},labelsep=period,name={Figure}}
\captionsetup[table]{labelfont={bf},labelformat={simple},labelsep=period,name={Table}}
\title{Decade-long timing of four GMRT discovered millisecond pulsars}
\author{
Shyam~S.~Sharma\altaffilmark{1}, 
Jayanta Roy\altaffilmark{1},
Bhaswati Bhattacharyya\altaffilmark{1} and Lina Levin\altaffilmark{2}}
\altaffiltext{1}{National Centre for Radio Astrophysics, Tata Institute of Fundamental Research, Pune 411007, India}
\altaffiltext{2}{Jodrell Bank Centre for Astrophysics, School of Physics and Astronomy, The University of Manchester, Manchester M13 9PL, UK}
\affil{}
%{~~~~~~~~~~~~~~~~~~~~~~~~~~~~~~~~~~~~~~~~~~~~~}\{ABSTRACT}

\section*{ABSTRACT}

%Recently, multiple pulsar timing arrays (PTAs) have discovered a pattern in the cross-correlation statistics of the millisecond pulsars' (MSPs) residuals that resembles the Hellings and Downs curve and suggests the possibility of an isotropic stochastic gravitational wave (GW) background. However, the S/N of the common correlated signal across pairs of MSPs given by each PTAs is less than 5, and only a high S/N cross-correlation curve will clarify the source of its origin. 
The discovery and timing follow-up of millisecond pulsars (MSPs) are necessary not just for their usefulness in Pulsar Timing Arrays (PTAs) but also for investigating their own intriguing properties. In this work, we provide the findings of the decade-long timing of the four MSPs discovered by the Giant Metre-wave Radio Telescope (GMRT), including their timing precision, model parameters, and newly detected proper motions. We compare the timing results for these MSPs before and after the GMRT upgrade in 2017, {characterise the improvement} in timing precision due to the bandwidth upgrade. We discuss the suitability of these four GMRT MSPs as well as the usefulness of their decade-long timing data for the PTA {experiments. It may aid} in the global effort to improve the signal-to-noise (S/N) of recently detected signature of gravitational waves in cross-correlation statistics of residuals of MSPs.

\section{Introduction}
\label{sec:intro}

Millisecond pulsars (MSPs) are fast rotating neutron stars that exhibit exceptional clock-like behaviour over time. Due to their rotational stability, we can probe the effects due to {the} interstellar medium along the line of sight as well as the effects that are intrinsic to them (e.g., \cite{1990ApJ...364..123F}). Pulsar timing (e.g., \cite{2004hpa..book.....L}) is a technique that utilises the time-of-arrival (ToA) information of the pulses coming from MSPs, {thus} revealing precise astrometric, rotational and binary models as well as propagation effects.

{The} long-term timing of MSPs helps to reduce covariance between system model parameters, resulting in more precise and accurate estimation of these parameters. It can reveal higher-order effects in the system, such as the higher order derivatives of spin frequency, temporal variations in orbital motion in the case of binaries, and a change of the electron-proton plasma along the line of sight (e.g., \cite{2004hpa..book.....L}). A better modelling of pulsars by taking these higher-order effects into account will result in more accurate predictions for their ToAs, making them better interstellar clocks.

\cite{1979ApJ...234.1100D} introduced the utilisation of long-term pulsar timing of MSPs to identify an isotropic stochastic gravitational wave (GW) background of cosmic origins. {Its} primary contributor is expected to be an ensemble of merging galaxies having supermassive black hole binaries at their centers \citep{2019A&ARv..27....5B}. A pulsar timing array (PTA) is an array of well-timed MSPs with varying angular separations in the sky designed to detect the effect of nano-Hertz (nHz) GWs on the cross-correlation of timing residuals from pairs of MSPs \citep{1983ApJ...265L..39H}. The current major PTAs are the North American Nanohertz Observatory for Gravitational Waves (NANOGrav) \citep{2009arXiv0909.1058J}, the Parkes Pulsar Timing Array (PPTA) \citep{2006ChJAS...6b.139M}, the European Pulsar Timing Array (EPTA) \citep{2006ChJAS...6b.298S}, the Chinese Pulsar Timing Array (CPTA) \citep{2016ASPC..502...19L}, the MeerKAT Pulsar Timing Array (MPTA) \citep{2016mks..confE..11B}, and the Indian Pulsar Timing Array (InPTA) \citep{2018JApA...39...51J}.

Recently, \cite{2023ApJ...951L...8A} (NANOGrav), \cite{2023ApJ...951L...6R} (PPTA), \cite{2023arXiv230616224A} {(EPTA $+$ InPTA)}, and \cite{2023RAA....23g5024X} (CPTA) have shown Hellings and Downs curve signatures in cross-correlation statistics of residuals for a set of MSPs. For the four PTAs, the signal-to-noise ratio (S/N) of the common correlated signal among the pairs of MSPs ranges from 3 to 5. S/N is defined here as $\rho/\sigma$ \citep{2005ApJ...625L.123J}. $\rho$ measures the similarity between the cross-correlation distribution of residuals (as a function of angular separation between MSPs) and the Hellings and Downs function \citep{1983ApJ...265L..39H}. $\sigma$ is the standard deviation of the cross-correlation distribution. According to \cite{2023ApJ...951L...8A}, the next step will be to merge data sets from NANOGrav, PPTA, EPTA, and InPTA, which will consist of around 80 MSPs and a time baseline of up to 24 years. This is expected to boost the detection significance of the common correlated signal. A high S/N cross-correlation curve will clarify the source of its origin, and a larger number of pulsars may reveal anisotropy in the GW background (e.g., \cite{2019OJAp....2E...8H}), polarisation structures (e.g., \cite{2022PhRvD.106b3004S}), and so on. Along with incorporating data sets from multiple PTAs, the individual PTAs intend to improve their sensitivity to this signal by increasing the number of MSPs and the observational timing baseline.

\cite{2013CQGra..30v4015S} describes the equation that shows the dependency of S/N, of the cross-correlation statistics, on the number of MSPs included in the GW detection experiment, the observational cadence, the timing baseline, the timing precision achieved for individual MSPs, and the amplitude of GWs. It shows that in the intermediate signal regime \citep{2013CQGra..30v4015S} the number of MSPs included in the experiment has the greatest influence on the detection significance of the common correlated GW signal, implying that identifying new MSPs which are suitable for PTAs should be of greater importance.

The Giant Meter-wave Radio telescope is an interferometer with 30 parabolic antennas and a longest baseline of 25 km. Each antenna dish has a 45 m diameter and two orthogonal polarisations\footnote{\url{http://www.gmrt.ncra.tifr.res.in}} \citep{1997hsra.book..217S}. The legacy GMRT system \citep{2010ExA....28...25R} is equipped with a GMRT Software Backend (GSB) with a maximum instantaneous observational bandwidth of 33 MHz within a frequency range of 120 to 1460 MHz. The upgraded GMRT (uGMRT; \cite{2017JAI.....641011R}, \cite{2017CSci..113..707G}) with GMRT Wide-band Backend (GWB) became operational in 2017, which can have a maximum instantaneous observational bandwidth of 400 MHz in the same frequency range as legacy GMRT. The GMRT High Resolution Southern Sky (GHRSS) survey [\cite{2016ApJ...817..130B}, \cite{2019ApJ...881...59B}, \cite{2022ApJ...934..138S}, \cite{2023ApJ...944...54S}, \cite{localisation_2023}, \cite{2023arXiv230701477S}] and \textit{Fermi}-directed survey [\cite{2013ApJ...773L..12B}, \cite{2015ApJ...800L..12R}, \cite{2021ApJ...910..160B}, \cite{2022ApJ...933..159B}] has so far discovered 12 MSPs using legacy GMRT and uGMRT systems. \iffalse Two of them are eclipsing binaries, while the remaining ten have not exibited any signature of eclipse yet.\fi For the timing campaign, we have been monitoring four of these regular MSPs that are bright in the lower frequency bands of the GMRT. These four MSPs were observed using the legacy GMRT system from 2011 to 2017, and after 2017, they have been observed using the uGMRT system. Phase-coherent timing has been obtained for the GSB observations for these 4 MSPs and was reported in \cite{2019ApJ...881...59B} and \cite{2022ApJ...933..159B}. \cite{2022ApJ...936...86S} reported the phase-coherent timing with GWB observations for the same MSPs, which accounts for 2$-$4 years of observational timing baseline. %This paper reports an increased timing baseline from the GWB observations of 3$-$5.5 years.

\iffalse
{In this study, we will present timing of GWB observations with an increased timing-baseline and combined it with legacy GMRT timing data sets, resulting in around a decade long baseline for the 4 MSPs. This long-term timing enabled us to precisely constraint the model parameters for these MSPs. We report the parameters that were detected for the first time for these MSPs. In addition, we report the change in line of sight electron density for these MSPs observed in GWB observations during the last 4$-$5 years. In this work, we performed a detailed comparison of the timing results from GSB, GWB, and GSB+GWB combined observations, and for the first time reporting the extent of improvement in pulsar timing observations after the GMRT upgrade. Finally, we comment on the suitability of these MSPs and usefulness of the decade long timing data for the PTAs.
\fi

In this paper, we  present the timing of GWB observations with an increased timing baseline and combine it with legacy GMRT timing data sets, resulting in around a decade-long baseline for the 4 MSPs. \iffalse We report the parameters that are detected for the first time for these MSPs. In this work, we perform a detailed comparison of the timing results from GSB, GWB, and GSB+GWB combined observations and, for the first time, report the extent of improvement in pulsar timing observations after the GMRT upgrade. Finally, we comment on the suitability of these MSPs and the usefulness of their decade-long timing data for the PTAs.\fi
Section \ref{sec:Section-2} and Section \ref{sec:Section-3} provides the specifics of the observation and the timing technique used in this study, respectively. Section \ref{sec:Section-4} reports the results from timing analysis, a comparison between GSB and GWB timing results, and a comparison of the timing residuals for the 4 GMRT MSPs with other PTAs. Section \ref{sec:Section-5} summarise the findings from the decade-long timing of these 4 MSPs and the suitability of GMRT discovered MSPs and their decade long timing data sets for the PTAs.

\section{Observation and data reduction}
\label{sec:Section-2}

For phased array (PA) GWB observations in uGMRT band-3 (300$-$500 MHz) and band-4 (550$-$750 MHz), 70$\%$ and 80$\%$ of the array is phased, providing gains of 7 K/Jy and 8 K/Jy, respectively. An array with similar gains as GWB is phased for PA beam observations in legacy GMRT band-3 (306$-$339 MHz) and band-4 (591$-$624 MHz) (\cite{2019ApJ...881...59B}). Table \ref{Observational_set_up} lists the configuration for the observations in various receiver backends, dedispersion modes, frequency bands, time resolution, and antenna count. The intra-channel smearing has been corrected for coherent dedispersion mode data sets with known DM values for the observed MSPs. The residual smearing for incoherent dedispersion mode data sets is determined by 2048 (GSB) or 4096 (GWB) channels over a bandwidth of 33 and 200 MHz, respectively.

\begin{table}[H]
    
    \centering
    \begin{adjustbox}{width=\columnwidth,center}
    \begin{tabular}{|c|c|c|c|c|c|c|c|}
    \hline
    
    GMRT &Mode & Frequency& Usable& Time& No. of\\
    Band &    & range (MHz) & bandwidth (MHz) & Resolution ($\mu s$) & Antennas\\
    \hline
    \hline
    GWB Band-3& I & 300-500 & 135 & 81.92 & 22\\
    GWB Band-4& I & 550-750 & 152 & 81.92 & 25\\
    \hline
    GWB Band-3 & C & 300-500 & 135 & 10.24/20.48/40.96$^{\dagger}$& 22\\
    GWB Band-4 & C & 550-750 & 152 & 10.24/20.48/40.96$^{\dagger}$ & 25\\
    \hline
    \hline
    GSB Band-3& I & 306-339 & 30 & 61.44 & 22\\
    GSB Band-4& I & 591-624 & 30 & 61.44 & 25\\
    \hline

    \end{tabular}
    \end{adjustbox}
    \caption{The table lists the details of the observational setup in different modes. C and I represent coherent and incoherent dispersion modes. In I mode, filterbank files have 2048 and 4096 channels in GSB and GWB backends, respectively.\\ $\dagger$ - In C mode, filterbank files have 512/1024/2048 channels in our observations. The table shows the time resolution corresponding to the filterbank with different numbers of channels.}
    \label{Observational_set_up}
\end{table}

We observed 4 GMRT-discovered MSPs, which are presented in Table \ref{Pulsars_table_observation}. The table shows the mean observation time, the GMRT backend utilised, the S/N in various frequency bands, the number of epochs in various observing modes and bands, and the timing baseline for the observed MSPs. The flux density values, profiles of these MSPs at various frequencies are reported in \cite{2022ApJ...936...86S}. The spin period and DM are listed in the timing ephemerides table (Table \ref{timing_solutions_full}) in Section \ref{sec:Section-4}.  

The four MSPs were selected from the set of GMRT discovered MSPs based on the S/N ratio (provided in the same table) and the timing precision reported in \cite{2022ApJ...936...86S}, \cite{2022ApJ...933..159B}, and \cite{2019ApJ...881...59B}. The four MSPs are bright in GMRT band-3 and band-4, with the exception of J1120$-$3618, which has significantly lower detection significance in band-4. These MSPs have been observed roughly with a monthly cadence for the last 7.7$–$11.0 years.

\begin{table}[H]
\centering

 \begin{tabular}{||c|c|c|c|c|c||}
 \hline
  & GMRT & Mean & Median S/N & No. of & Timing \\ 
 MSP & Backend & Observation &  & Epochs & baseline \\
 & & time &  Band 3~~~Band 4 &  Band 3~~~Band 4 &(years)\\
  &  & ({min})&   &  C/I~~~~~~~~~~C/I & \\
 \hline\hline
 J1120$-$3618 & GWB &52 & 64~~~~~~~~~~-~ & 19/13~~~~~~~~~ -/- ~& 4.2\\
              & GSB &   & \iffalse 26~~~~~~~~~~-~ \fi & - /26~~~~~~~~~ -/- ~& 4.3\\
\hline
 J1646$-$2142 & GWB & 43 & 70~~~~~~~~~45 & 26/17~~~~~~~13/10 & 4.0\\
              & GSB &    & \iffalse 33~~~~~~~~~21 \fi & - /58~~~~~~~~-/10 & 5.8\\
\hline
 J1828$+$0625 & GWB & 46 & 35~~~~~~~~~26 & 19/13~~~~~~~~~4/4 & 3.1\\
              & GSB &    &\iffalse 13~~~~~~~~~10 \fi & - /55~~~~~~~~~-/2 & 6.1\\
\hline
 J2144$-$5237 & GWB & 55 & 48~~~~~~~~~58 & 27/24~~~~~~~~~4/9 & 5.5\\
              & GSB &    & \iffalse 19~~~~~~~~~24 \fi & ~- /24~~~~~~~~~ -/- ~& 1.8\\
 \hline
\end{tabular}
\caption{Observational parameters of the four MSPs used in this work. The fifth column lists the number of observations collected for each MSP. It should be noted that the majority of the C and I observations indicated in the table's fifth column were recorded simultaneously. The observation duration for MSPs with GSB and GWB backend was roughly the same.}
\label{Pulsars_table_observation}
\end{table}

Following the observations with the GWB, the GMRT raw PA beam data sets are converted to \texttt{filterbank} format, which is then incoherently dedispersed and folded with the \texttt{PREPFOLD} command of \texttt{PRESTO} \citep{2011ascl.soft07017R}. For folding purposes, we make use of the ephimeris presented in \cite{2022ApJ...933..159B}, and \cite{2019ApJ...881...59B}. The folded data cubes (\texttt{PFDs}) produced by the \texttt{PREPFOLD} command are then converted to \texttt{FITS} format using the \texttt{PAM} command of \texttt{PSRCHIVE} \citep{2012AR&T....9..237V}. Following the same profile bin configuration as \cite{2022ApJ...936...86S}, we created 128 and 64 bins for the coherent and incoherent dedispersed profile, respectively, in \texttt{FITS} files for all the MSPs. The \texttt{FITS} files are averaged to a single time integration, 64/128 bins, and 16 frequency subbands. These \texttt{FITS} files are then used for further timing analysis described in Section-\ref{sec:Section-3}.

For all GSB observations, phase coherent timing has been established in \cite{2022ApJ...933..159B}, and \cite{2019ApJ...881...59B}. The GSB ToA from these works are being used directly for this study for aiding long-term timing.

\section{Timing technique}
\label{sec:Section-3}

This section outlines the procedure that we used to create templates for the MSPs, derive DM from each epoch observation, and produce ToAs for the timing analysis.

Procedure for creating templates: We collected all high S/N \texttt{FITS} files in a specified frequency band/mode for a particular MSP. We use \texttt{ppalign} module of ``\texttt{PulsePortraiture}"\footnote{\label{pulse_portraiture}\url{https://github.com/pennucci/PulsePortraiture}} (\cite{2019ApJ...871...34P}, \cite{2016ascl.soft06013P}, and \cite{2014ApJ...790...93P}) software to create a 2-dimensional template using the FITS files. The \texttt{ppalign} module {aligns} the selected set of \texttt{FITS} files iteratively. This module is equipped with a robust alignment algorithm that takes into consideration not only a constant temporal phase offset between different epoch observations but also rotates each frequency subband profile of an epoch by an offset proportional to $\nu^{-2}$. Here, $\nu$ is the frequency of the corresponding frequency subband profile. This results in an aligned set of FITS files which are then averaged while keeping the frequency resolution to obtain a 2-dimensional template (function of frequency and phase bins). 

For each MSP the resultant 2-dimensional template is having one time integration, 64/128 profile bins, and 16 freq-subbands. The 2-dimensional template is then frequency averaged to provide a single profile with 64/128 bins. This profile is then used as a template for the particular dedispersion mode and frequency band for that MSP.  Note that we use separate templates for the two dedispersion modes and different uGMRT receiver bands for each MSP. 

Deriving DM: We considered all the \texttt{FITS} files of a single MSP in a single band and a single dedispersion mode at a time. The template corresponding to this band, mode and MSP is used to generate freq-subband ToAs from the \texttt{FITS} files (having 1 time-integration, 64/128 bins, 16 frequency-subbands), resulting in 16 subband ToAs per epoch. We use the \texttt{PAT} command of \texttt{PSRCHIVE} to generate ToAs. \texttt{TEMPO2} \citep{2012ascl.soft10015H} is used to fit for the DM using these subband ToAs for each epoch, keeping the other parameters fixed while fitting.

ToAs for timing: We averaged all \texttt{FITS} files corresponding to a specific MSP, band, and mode in frequency so that they have 1 time-integration, 64/128 bins, and 1 frequency-subband. The corresponding template for this set is used to generate a single ToA for each epoch observation. Again, we create ToAs using the \texttt{PAT} command of \texttt{PSRCHIVE}. These band-averaged ToAs are used for timing purposes.

\label{sec:Section-3}

\section{Timing Results}
\label{sec:Section-4}

We have measured the temporal variation of DM for each of the four MSPs from the band-3 GWB data sets, and show them in each pulsar subsection below. %Figures \ref{J1120_DM}, \ref{J1646_DM}, \ref{J1828_DM}, \ref{J2144_DM} show the temporal variation of DM for the four MSPs from the band-3 GWB data sets.
We did not include the DM values from GWB band-4 and the GSB data sets in these plots since their error bars (GWB band-4: $\sim 5\times 10^{-3}$ pc cm$^{-3}$; GSB band-3/4: $\sim 10^{-1}$ pc cm$^{-3}$ or higher) are significantly larger than those from the GWB band-3 data sets  ($\sim 5\times 10^{-4}$ pc cm$^{-3}$). Due to the very large DM error bars in the GSB data sets, we did not consider the DM variations with time for any of the MSPs to treat GSB and GWB data sets equally when doing timing. Instead, we used a global DM for the combined timing of band-3 + band-4 and GSB+GWB data sets (named GSB+GWB from now on) of each MSP.  

Table \ref{Median_ToA} shows the number of ToAs corresponding to individual frequency bands and receiver backends for each MSP, as well as the median ToA uncertainty and the root-mean-square (RMS) of post-fit timing residuals obtained from the individual GSB, GWB, and GSB+GWB timing for these MSPs. 
%Figures \ref{J1120_timing}, \ref{J1646_timing}, \ref{J1828_timing}, \ref{J2144_timing} show 
We show the post-fit residuals for each MSPs, corresponding to GSB+GWB timing, in each pulsar subsection below. Table \ref{timing_solutions_full} shows the timing ephimerides and derived parameters obtained from phase-coherent timing of GSB+GWB. 

\subsection{J1120$-$3618}

J1120$-$3618 is a binary MSP with a spin-period of $\sim$5.56 ms, an orbital period of $\sim$5.66 days, and a DM of $\sim$45.12 $pc\,cm^{-3}$. \iffalse For this MSP, the median ToA error bar in GWB band-3 is 6.74 $\mu s$, which is $\sim$3 times smaller than its median GSB ToA error.\fi The timing precision obtained in individual GWB and GSB timing is 6.1 and 27.0 $\mu s$ respectively, each spanning more than 4 years of baseline. We achieved an overall timing precision of 10.5 $\mu s$ after combining the GSB and GWB timing data sets (spanning 10.3 years), which is clearly limited by the larger error bars of ToAs in GSB timing data. 

This MSP's line of sight is crossing an electron-rich medium for the last 4–5 years, as indicated by its DM trend seen in GWB band-3. Its DM value has increased by 8.6$\times$10$^{-3}$ $pc\,cm^{-3}$ from February 2018 to January 2023 while the median DM error bar for this MSP in GWB band-3 is 1.0$\times$10$^{-3}$ $pc\,cm^{-3}$. 

\begin{figure}[H]
\centering
        \includegraphics[width=0.9\linewidth,keepaspectratio]{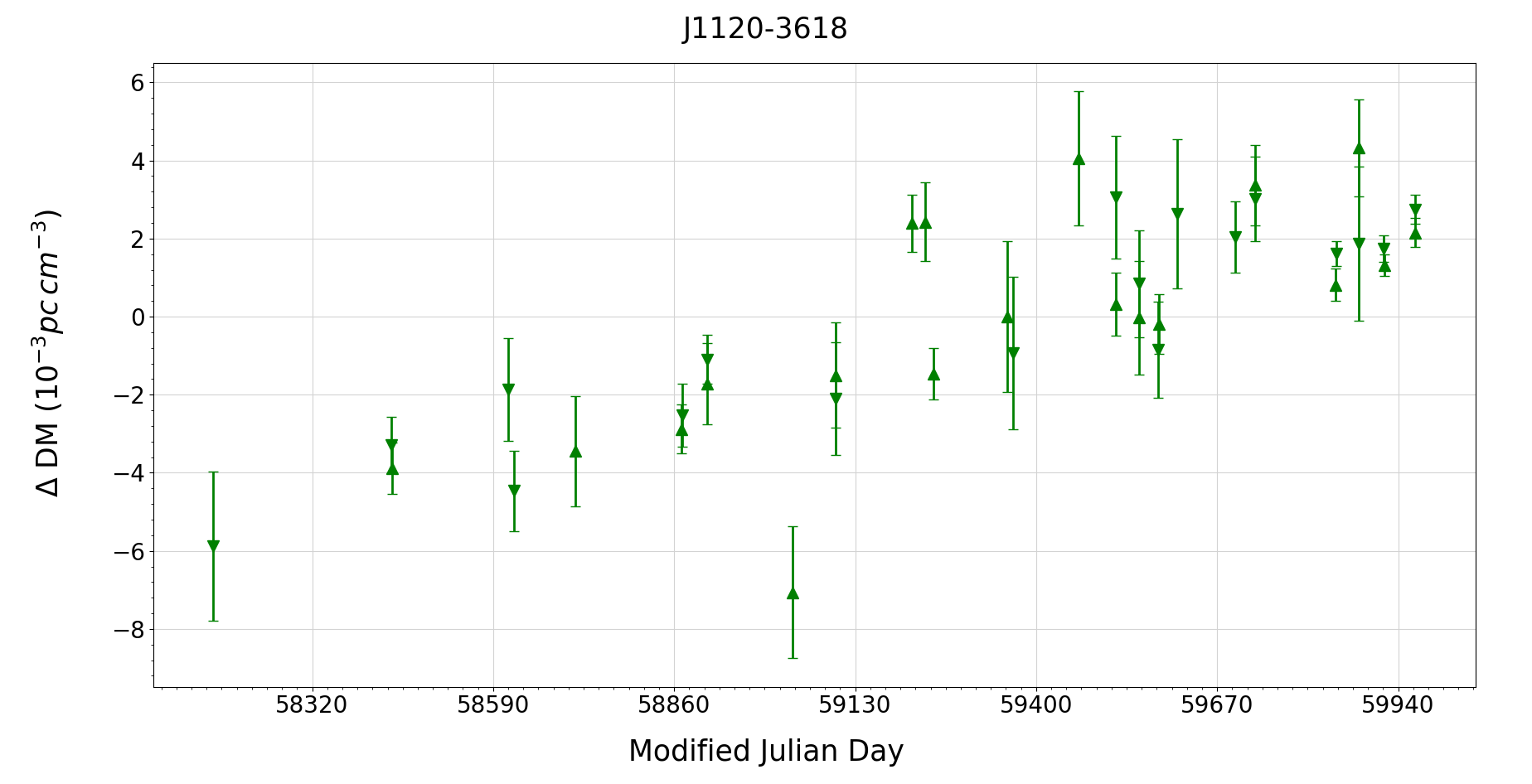}
        \vspace{-5mm}
    \caption{Figure showing DM variation with time for the MSP J1120$-$3618 in GWB band-3 of the uGMRT. Up and down triangles represent coherently and incoherently dedispersed data, respectively.}
    \label{J1120_DM}
\end{figure}
\begin{figure}[H]
\centering
        \includegraphics[width=0.9\linewidth,keepaspectratio]{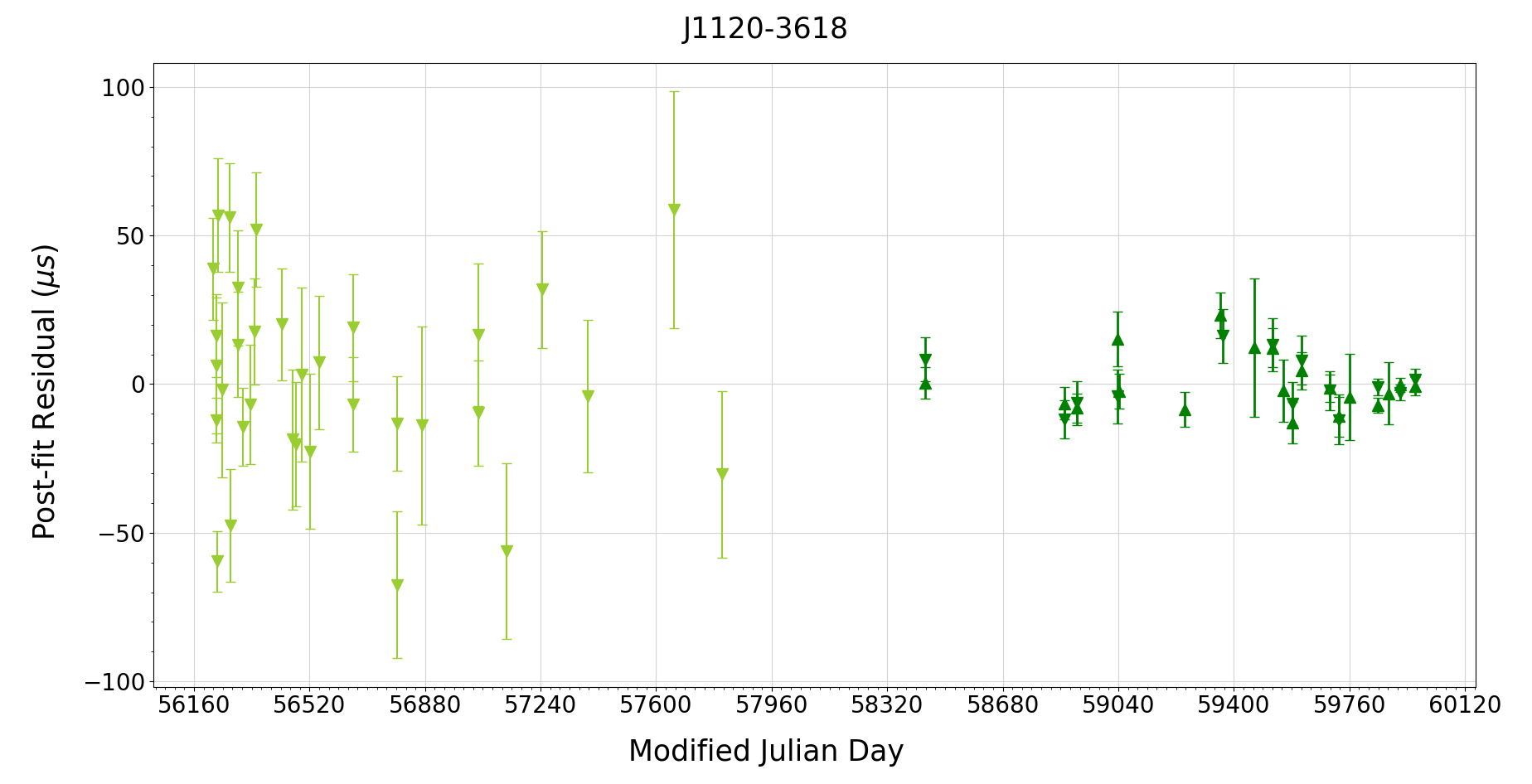}
        \vspace{-5mm}
    \caption{Post-fit timing residuals vs MJD for MSP J1120$-$3618. Yellow green and green colors are used for GSB band-3 and GWB band-3, respectively. Up and down triangles represent coherently and incoherently de-dedispersed data sets, respectively.}
    \label{J1120_timing}
\end{figure}

%The proper motion for this MSP is expected given the change in DM values supporting the change of line of sight electron density. 
We measure a total proper motion for this MSP, for the first time, of 6.0(6) $mas/yr$ from GSB+GWB timing, and individually 5.2(6) $mas/yr$ in right ascension (RA) and 2.8(7) $mas/yr$ in declination (DEC). The proper motion for J1120$-$3618 falls within the typical proper motion range for known MSPs [few tens of $mas/yr$; ATNF pulsar catalogue\footnote{https://www.atnf.csiro.au/research/pulsar/psrcat/} \citep{2005AJ....129.1993M}].

{We estimate the distance\footnote{https://pulsar.cgca-hub.org/compute} to this MSP using its DM, position, and the galactic electron density models NE2001 \citep{2002astro.ph..7156C} and YMW16 \citep{2017ApJ...835...29Y}, which result in a distance of 1.75 and 0.95 $kpc$, respectively. Using these estimates and the proper motion value for this MSP, we calculate transverse velocities of 49 $km/s$ (for 1.74 $kpc$; NE2001) and 27 $km/s$ (for 0.95 $kpc$; YMW16). These transverse velocity values are significantly lower than the maximum expected space velocity of the MSPs ($\sim$270 $km/s$) in the scenario of spherically symmetric supernova explosions, suggesting that this MSP did not experience an asymmetric kick during its recoiling phase \citep{1996A&A...315..432T}.
}

\iffalse
{Estimate the 2D transverse velocity (VT) using the relation VT $=$ 4.74 $\mu$T (mas yr$^{-1}$) d (kpc), where d is the distance to the pulsar. Using two galactic electron density models, NE2001 (Cordes $\&$ Lazio 2002) and YMW16 (Yao et al. 2017), the DM distance (d) needs to be estimated which will give two values of VT. Compare the VT value with values predicted for different evolutionary mechanisms for MSP formations, Tauris $\&$ Bailes (1996) predicted mean recoil velocity for MSPs to be 110 km/s assuming symmetric supernova explosion and 160 km/s with the inclusion of random asymmetric kicks. Then calculate the intrinsic spin period derivative using Equation (2) of Toscano et al. (1999), considering VT value given by YWM16 model. } \fi\iffalse \cite{2022ApJ...933..159B} reported the eccentricity for this MSP which is found by timing only the GSB data; addition of GWB timing data to it resulted in a value for eccentricity with large error bars, therefore we also excluded its fitting in timing.\fi 
The timing-fit for J1120$-$3618 yields a spin-period derivative (P1) of 9.4(3) $\times$ 10$^{-22}$ s/s, an order of magnitude less than the other three MSPs in this work. {It has an intrinsic spin-period derivative \citep{1999MNRAS.307..925T} of $\sim$5.5 $\times$ 10$^{-22}$ s/s ($\sim$0.6 P1) [determined using Equation 2 of \cite{1999MNRAS.307..925T}, proper motion, and the distance measurements from the YMW16 model].} The P1-value of J1120$-$3618 is the fifth smallest of all MSPs currently known.

\begin{figure}[H]
\centering
        \includegraphics[width=0.49\linewidth,keepaspectratio]{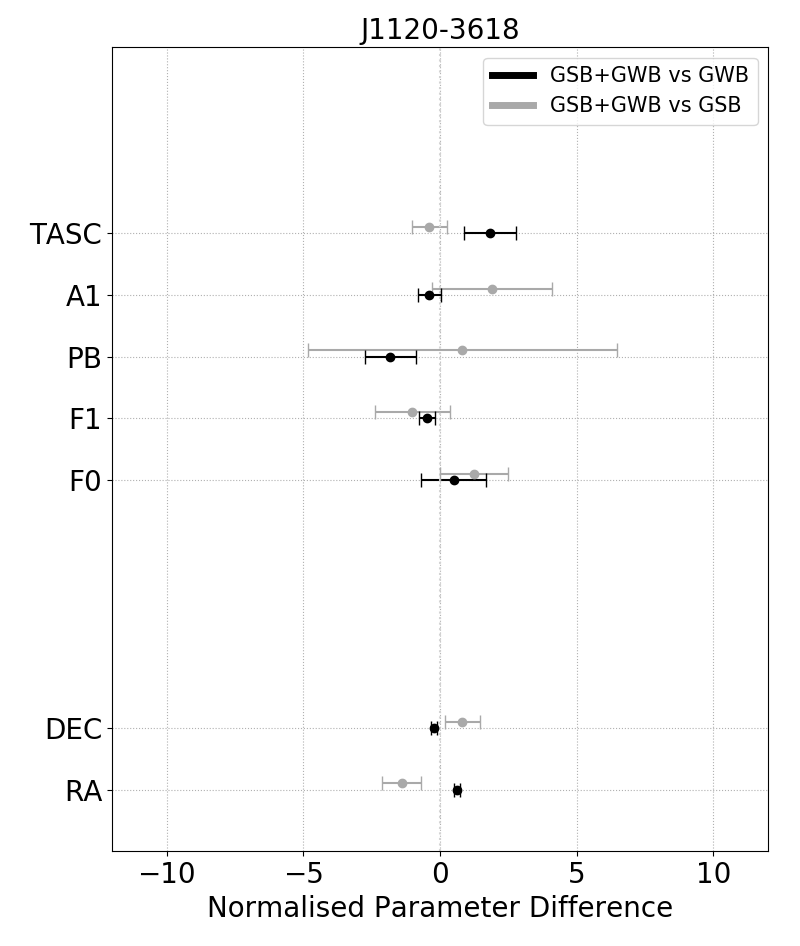}
        
    \caption{Figures showing a comparison of our fitted timing models for J1120$-$3618. The y-axis shows the name of the fitted parameter.\\
    Black color points/error-bars: The x-axis shows the differences of the fitted parameter (astrometric, spin, and binary) values between GSB+GWB timing and only GWB timing normalized by the uncertainties of GWB, i.e., $(X_{GSB+GWB}-X_{GWB})/\sigma_X^{GWB}$ where X is the MSP's model parameter. The error bars have a length equal to the ratio of parameter uncertainties from GWB and GSB+GWB models, i.e., $\sigma_X^{GWB}/\sigma_X^{GSB+GWB}$.\\
    Dark gray color points/error-bars: The x-axis shows the difference of parameter values between GSB+GWB timing and only GSB timing normalized by the uncertainties of GSB, i.e.,  $(X_{GSB+GWB}-X_{GSB})/\sigma_X^{GSB}$. The error bar length in this case is $\sigma_X^{GSB}/\sigma_X^{GSB+GWB}$.}
    
    \label{J1120_model_plot}
\end{figure}

\iffalse
The post-fit residuals for this MSP show a systematic bump in GWB timing data that appears like a sinusoidal. The cause of this change in the pattern of timing residuals is currently unclear but we anticipate finding it out after adding a few more years of GWB data.
\fi
Figure \ref{J1120_model_plot} shows a comparison of fitted model parameters from GWB, GSB, and GSB+GWB timing for this MSP. Commonly fitted model parameters (listed in Figure \ref{J1120_model_plot}) obtained from GSB+GWB timing are on average 1.1 and 3.6 times more precise than individual GWB and GSB timing, respectively. The model parameters from GSB+GWB timing are consistent with individual GWB or GSB timing within $\pm 1 \sigma$, where $\sigma$ is the error bar of the fitted model parameter from either GWB or GSB. 

%%%%%%%%%%%%%%%%%%%%%%%%%%%%%%%%
\subsection{J1646$-$2142}

J1646$-$2142 is an isolated MSP with a spin-period of $\sim$5.85 ms and a DM of $\sim$29.74 $pc\, cm^{-3}$. \iffalse The median ToA errors in GWB band-3 and band-4 for this MSP are 4.09 and 5.97 $\mu s$, respectively, which are $\sim$4 and $\sim$3 times smaller than the median ToA errors in GSB band-3 and band-4. \fi Timing precision in GWB and GSB timing are 7.3 and 15.7 $\mu s$, respectively. With combined GSB and GWB timing data, we achieved a timing precision of 11.1 $\mu s$ for a span of 11.0 years. 

The median DM uncertainty for this MSP in GWB band-3 and band-4 is 6.5$\times$10$^{-4}$ and 5.1 $\times$10$^{-3}$ $\,pc\,cm^{-3}$, respectively. We see an overall DM change of less than $\pm$3 $\sigma_{DM}$ from July 2017 to October 2022, where $\sigma_{DM}$ is the median DM uncertainty in GWB band-3.

\begin{figure}[H]
\centering
        \includegraphics[width=0.9\linewidth,keepaspectratio]{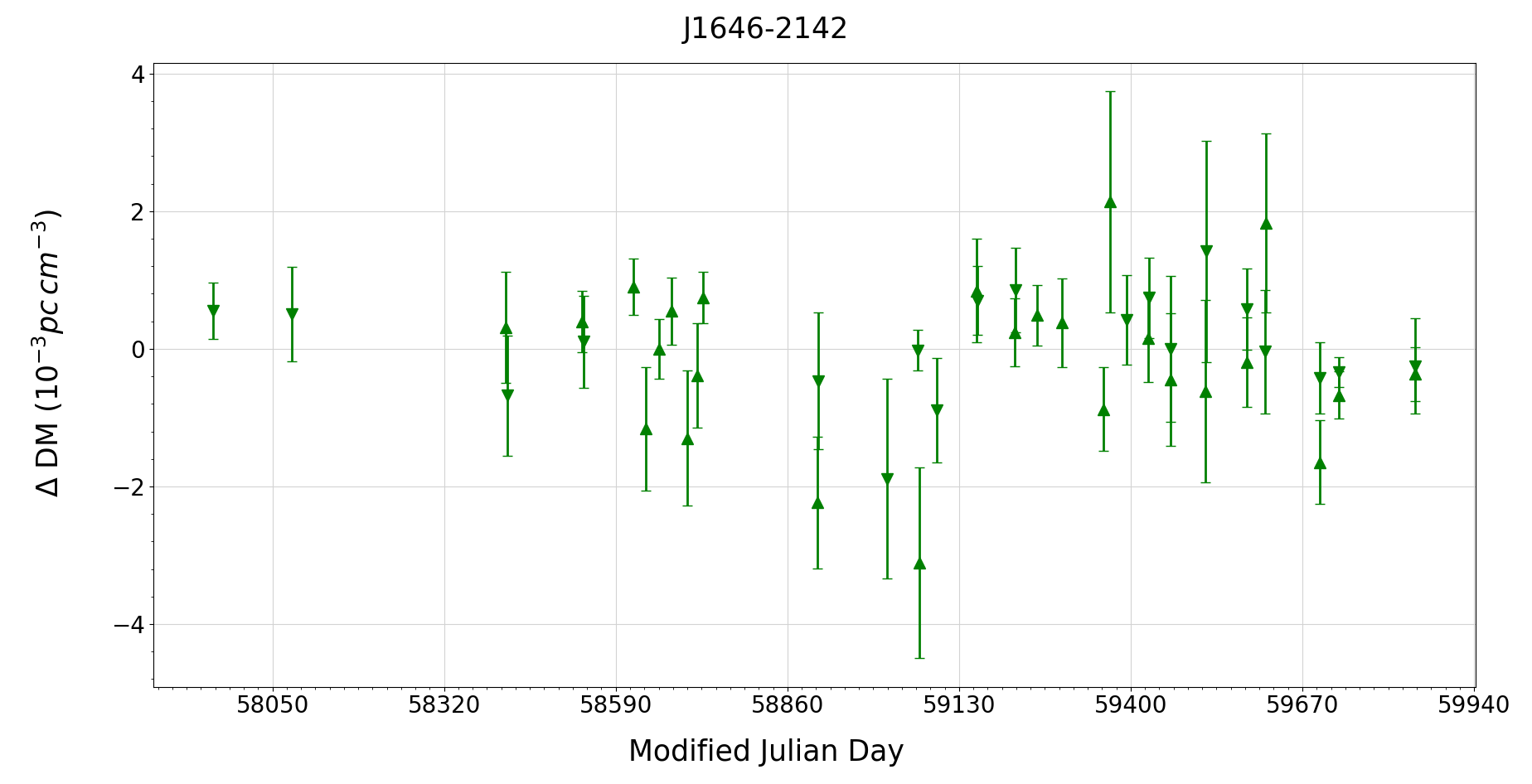}
        \vspace{-5mm}
    \caption{Figure showing DM variation with time for the MSP J1646$-$2142 in GWB band-3 of the uGMRT. The rest of the plotting approach is the same as in Figure \ref{J1120_DM}.}
    \label{J1646_DM}
\end{figure}
\begin{figure}[H]
\centering
        \includegraphics[width=0.9\linewidth,keepaspectratio]{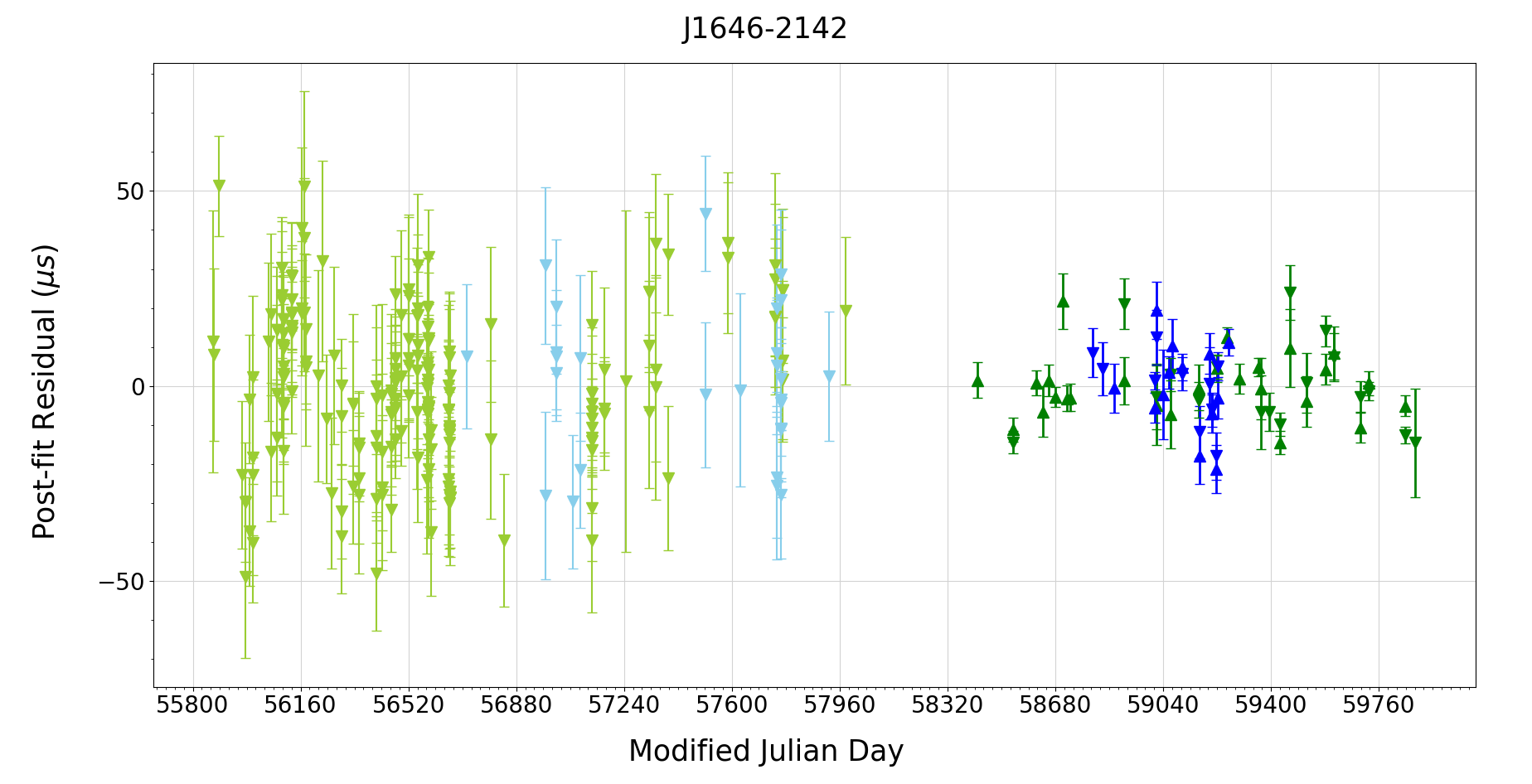}
        \vspace{-5mm}
    \caption{Post-fit timing residuals vs MJD for MSP J1646$-$2142. Yellow green, green, sky blue, and blue colors are used for GSB band-3, GWB band-3, GSB band-4, and GWB band-4, respectively. Up and down triangles represent coherently and incoherently de-dedispersed data sets, respectively.}
    \label{J1646_timing}
\end{figure}

\begin{figure}[H]
\centering
        \includegraphics[width=0.49\linewidth,keepaspectratio]{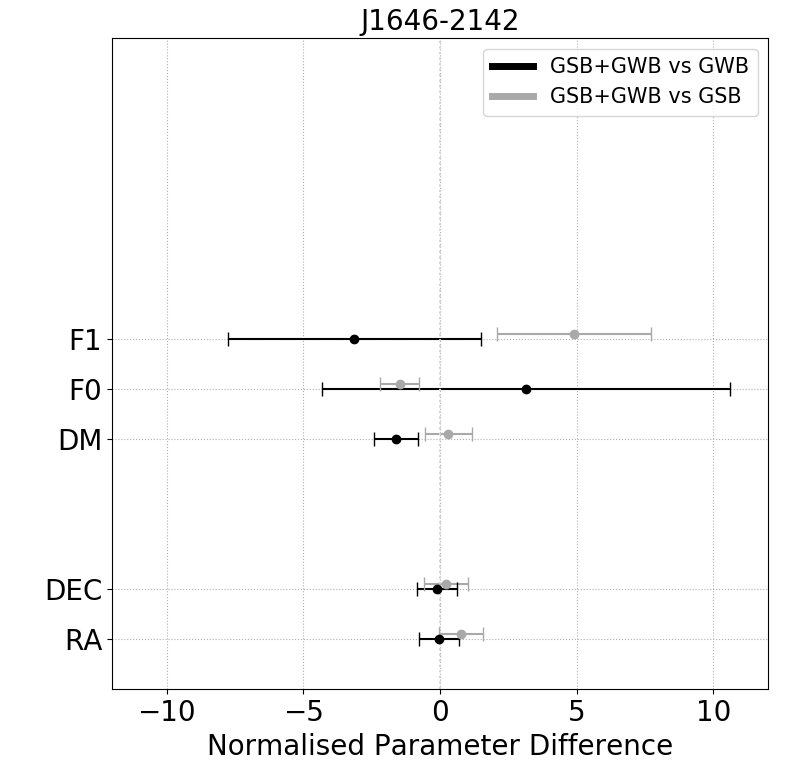}
        
    \caption{Figure showing a comparison of the fitted timing models for J1646$-$2142 in GSB+GWB timing to individual GWB and GSB timing. The plotting style is the same to Figure  \ref{J1120_model_plot}.}
    
    \label{J1646_model_plot}
\end{figure}

The GSB+GWB timing fit resulted in insignificant detection of proper motion; therefore, we excluded its fitting in timing. Figure \ref{J1646_model_plot} shows differences in model parameters and their precision in GSB+GWB timing against individual GWB and GSB timing. We see an average improvement in model parameter precision of 5.7 and  2.4 times in GSB+GWB timing when compared individually to GWB and GSB timing, respectively. The model parameters from GSB+GWB timing are consistent with individual GSB or GWB timing within $\pm 1 \sigma$, where $\sigma$ is the error bar of the fitted model parameter from either GSB or GWB.

%%%%%%%%%%%%%%%%%%%%%%%%%%

\subsection{J1828$+$0625}

J1828$+$0625 is a binary MSP with a spin-period of $\sim$3.63 ms, an orbital period of $\sim$77.92 days, and a DM of $\sim$22.42 $pc\, cm^{-3}$. \iffalse The median ToA error for this MSP in GWB band-3 and band-4 is 4.51 and 4.18 $\mu s$, respectively, which is $\sim$5 and $\sim$7 times smaller than the median ToA errors in GSB band-3 and band-4.\fi We achieved timing precision of 6.2 and 20.9 $\mu s$ in individual timing of the GWB and GSB data sets, respectively. The combined GSB+GWB timing resulted in a timing precision of around 11.8 $\mu s$ for a span of 10.9 years.

The median DM uncertainty of this MSP in GWB band-3 and band-4 is 8.1$\times$10$^{-4}$ and 5.2$\times$10$^{-3}$ $pc\,cm^{-3}$, respectively. We find that the DM variation for this MSP is confined within $\pm$3 $\sigma_{DM}$ from June 2019 to September 2022, where $\sigma_{DM}$ is the median DM uncertainty in GWB band-3. 

\begin{figure}[H]
\centering
        \includegraphics[width=0.9\linewidth,keepaspectratio]{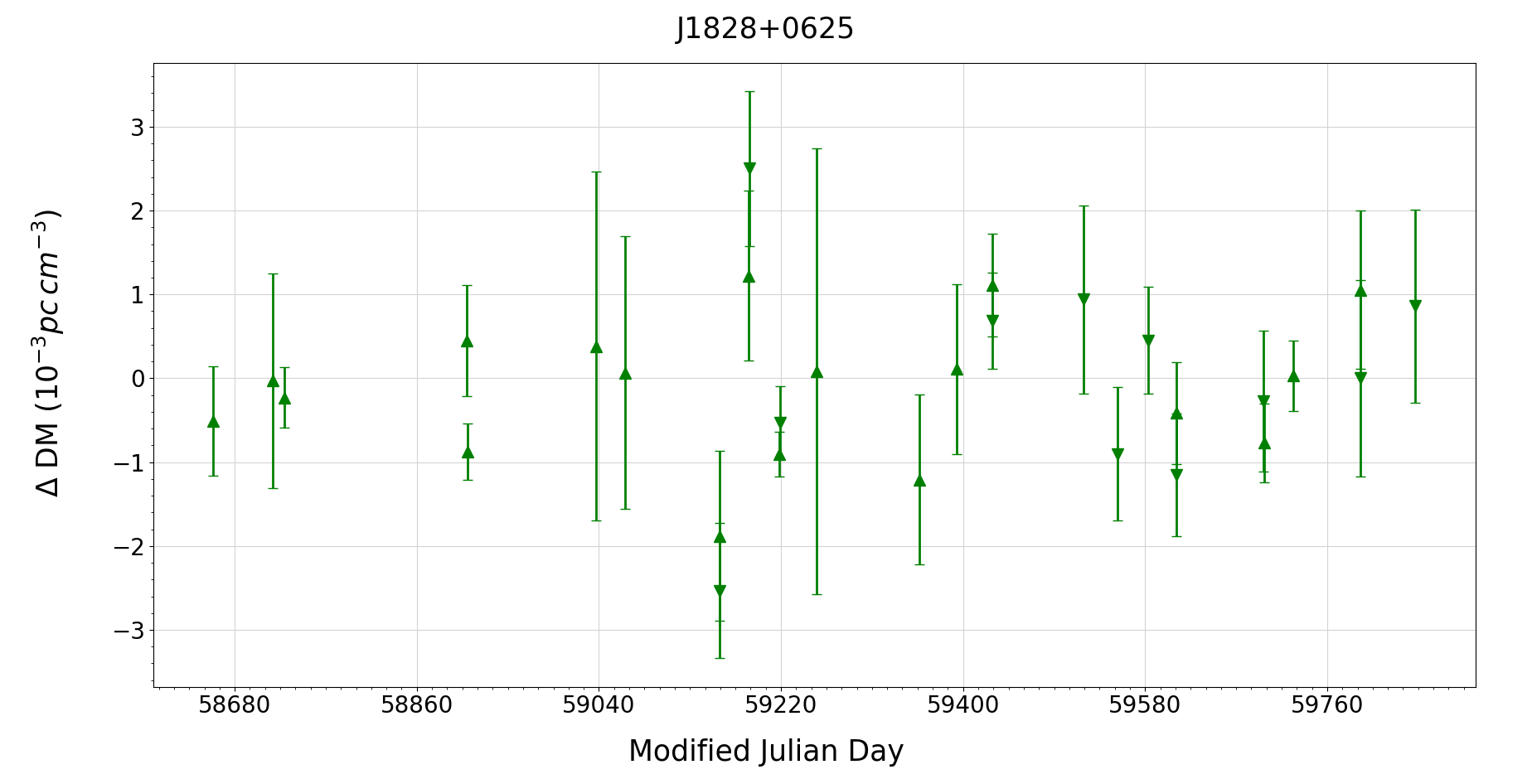}
        \vspace{-5mm}
    \caption{Figure showing DM variation with time for the MSP J1828$+$0625 in GWB band-3 of the uGMRT. The rest of the plotting approach is the same as in Figure \ref{J1120_DM}.}
    \label{J1828_DM}
\end{figure}
\begin{figure}[H]
\centering
        \includegraphics[width=0.9\linewidth,keepaspectratio]{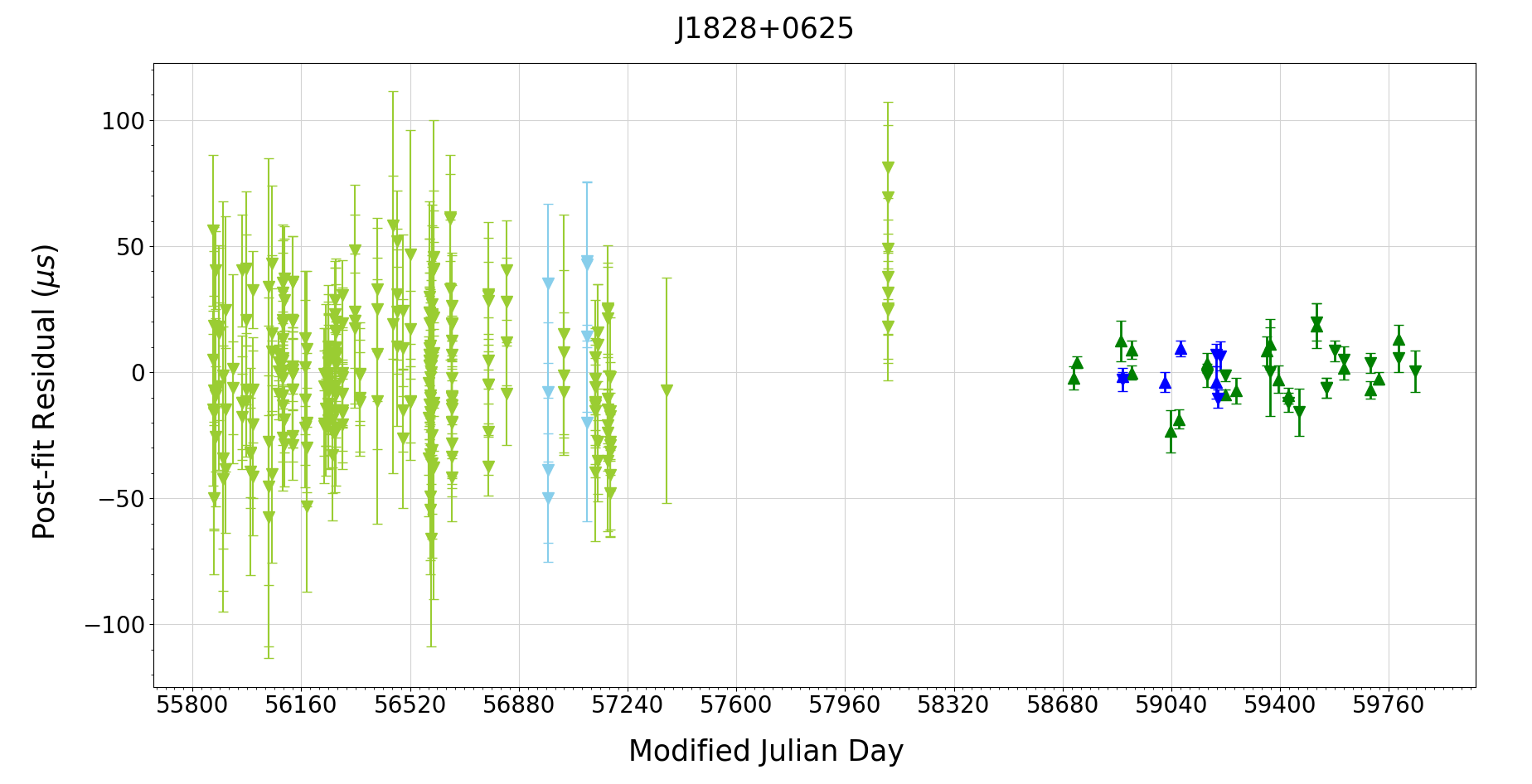}
        \vspace{-5mm}
    \caption{Post-fit timing residuals vs MJD for MSP J1828$+$0625. The rest of the plotting approach is the same as in Figure \ref{J1646_timing}.}
    \label{J1828_timing}
\end{figure}

For this MSP, we estimated a total proper motion of 14.4(2) $mas/yr$, which is well within $\pm$2$\sigma$ of the values reported in  \cite{2022ApJ...933..159B}.
%While the proper motion in RA reported in \cite{2022ApJ...933..159B} [-15.3(11) $mas/yr$] and in this study [-14.0(2) $mas/yr$] agrees well, the proper motion value in DEC in their studies [-5.8(15) $mas/yr$] is noticeably different from the value in this study [-3.0(2) $mas/yr$].
The increased timing span with precise GWB ToAs allows to derive the proper motion with at least 5 times higher precision than the values reported in \cite{2022ApJ...933..159B} using only GSB data.

{The distance estimates for this MSP using the NE2001 and YMW16 models are 1.12 and 1.00 $kpc$, respectively. Proper motion and these distances give a transverse velocity of 77 km/s (for 1.12 $kpc$; NE2001) and 68 km/s (for 1.00 $kpc$; YMW16) for this MSP, which is comparable to the transverse velocities of J1120$–$3618 and suggests the absence of an asymmetric kick during recoiling. The timing fit for this MSP resulted in a precise P1 value of 4.6882(9) $\times$ 10$^{-21}$ s/s. Using its transverse velocity and distance (for YMW16), the intrinsic spin period derivative is estimated to be $\sim$2.7 $\times$ 10$^{-21}$ s/s ($\sim$0.6 P1).}

%{Similar to J1120, estimate VT, compare with theoretical model prediction, estimate the intrinsic P1. Add VT, intrinsic P1 in the Table 4 in the derived parameters.}

\begin{figure}[H]
\centering
        \includegraphics[width=0.49\linewidth,keepaspectratio]{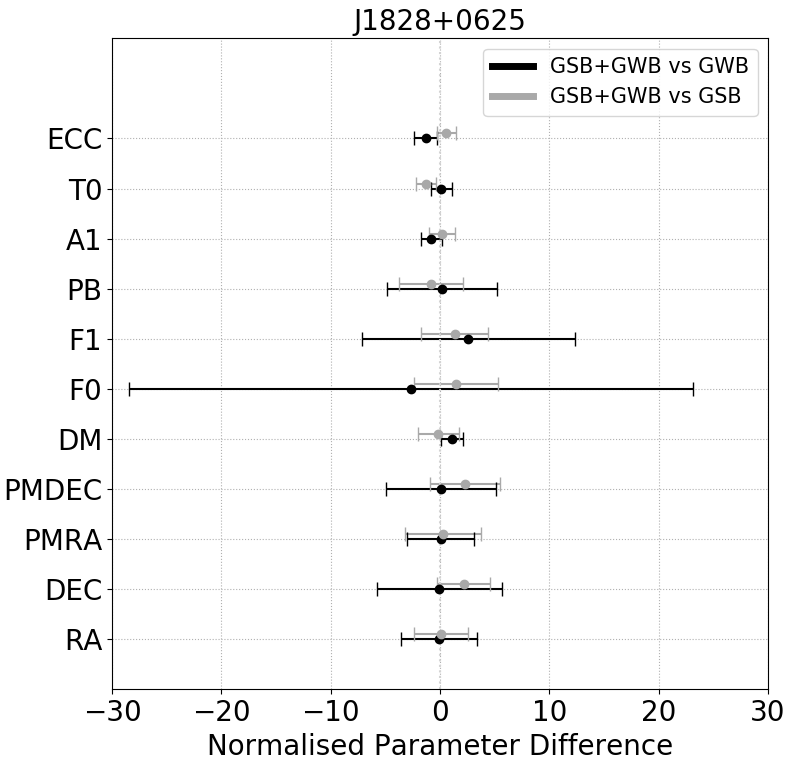}
        
    \caption{Figure showing a comparison of the fitted timing models for J1828$+$0625 in GSB+GWB timing to individual GWB and GSB timing. The plotting style is the same to Figure  \ref{J1120_model_plot}.}
    
    \label{J1828_model_plot}
\end{figure}

Figure \ref{J1828_model_plot} compares model parameters obtained from GSB+GWB timing to individual GSB or GWB timing. The model parameters obtained from GSB+GWB timing are on average 10.5 times more precise than individual GWB timing while being 4.5 times more precise than individual GSB timing. The model parameters resulting from GSB+GWB timing are in agreement with the parameters obtained from individual GWB or GSB timing within $\pm$1 $\sigma$, where $\sigma$ is the error in model parameter obtained from individual timing of GWB or GSB data sets.

%%%%%%%%%%%%%%%%%%%%%%%%%%%%%%

\subsection{J2144$-$5237}

J2144$-$5237 is a binary MSP with a spin-period of $\sim$5.04 ms, an orbital period of $\sim$10.58 days, and a DM of $\sim$19.55 $pc\, cm^{-3}$. \iffalse The median ToA error in GWB band-3 is 3.28 $\mu s$, which is 6 times smaller than the median ToA error in GSB band-3. The median ToA error in GWB band-4 is 4.32 $\mu s$, while GSB band-4 ToAs were not available for this MSP.\fi The timing precision achieved in individual timing of GWB and GSB data is 9.0 and 17.1 $\mu s$, respectively, whereas combined GSB+GWB timing resulted in a timing precision of 10.6 $\mu s$ over a 7.7-year span. 

The median DM uncertainty for this MSP in GWB band-3 and band-4 is 4.3$\times$10$^{-4}$ and 5.8$\times$10$^{-3}$ $pc\,cm^{-3}$, respectively. We see small variations in DM from July 2017 to January 2023, which are contained within $\pm$3 $\sigma_{DM}$, where $\sigma_{DM}$ is the DM uncertainty in GWB band-3.  
\begin{figure}[H]
\centering
        \includegraphics[width=0.9\linewidth,keepaspectratio]{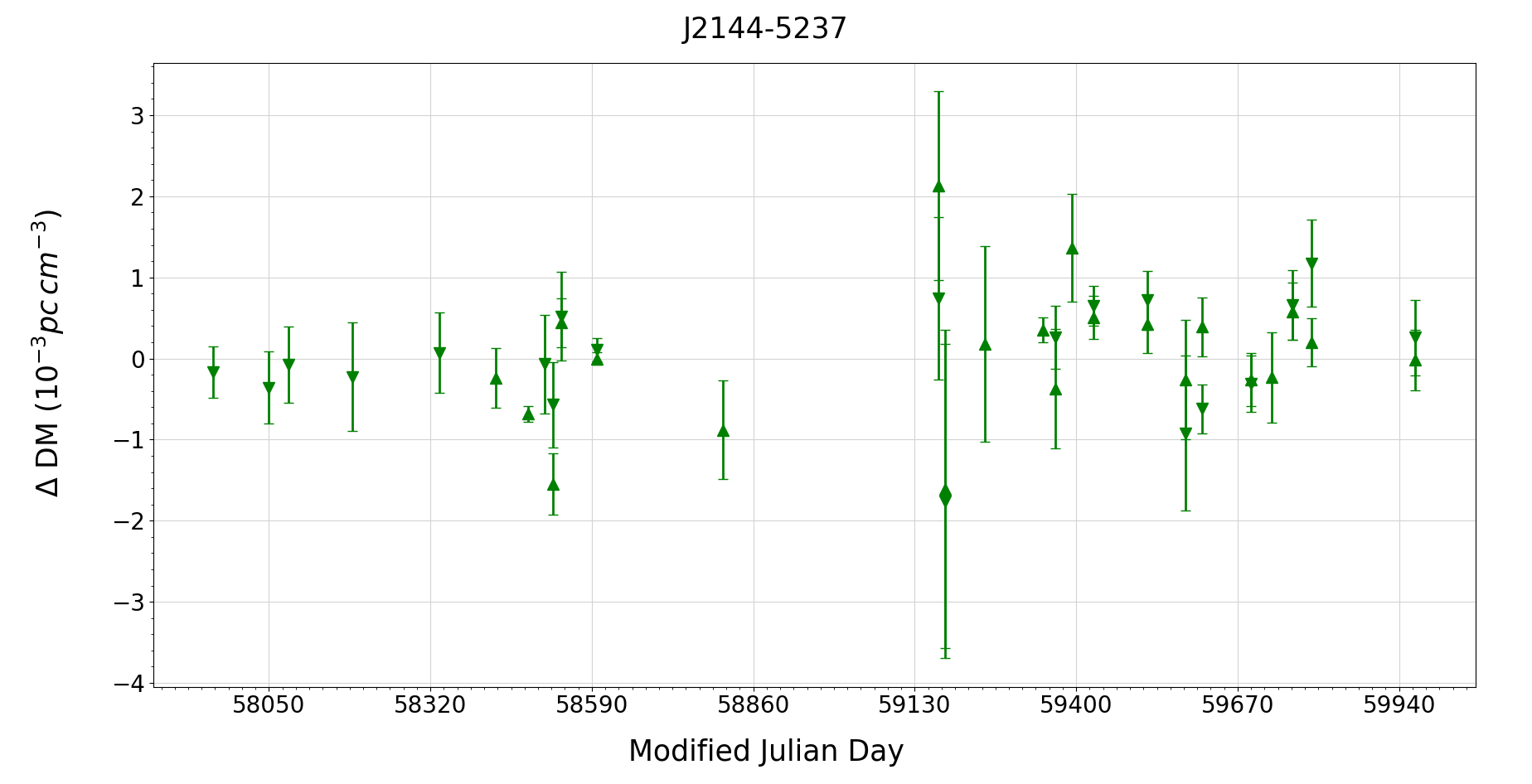}
        \vspace{-5mm}
    \caption{Figure showing DM variation with time for the MSP J2144$-$5237 in GWB band-3 of the uGMRT. The rest of the plotting approach is the same as in Figure \ref{J1120_DM}. }
    \label{J2144_DM}
\end{figure}
\begin{figure}[H]
\centering
        \includegraphics[width=0.9\linewidth,keepaspectratio]{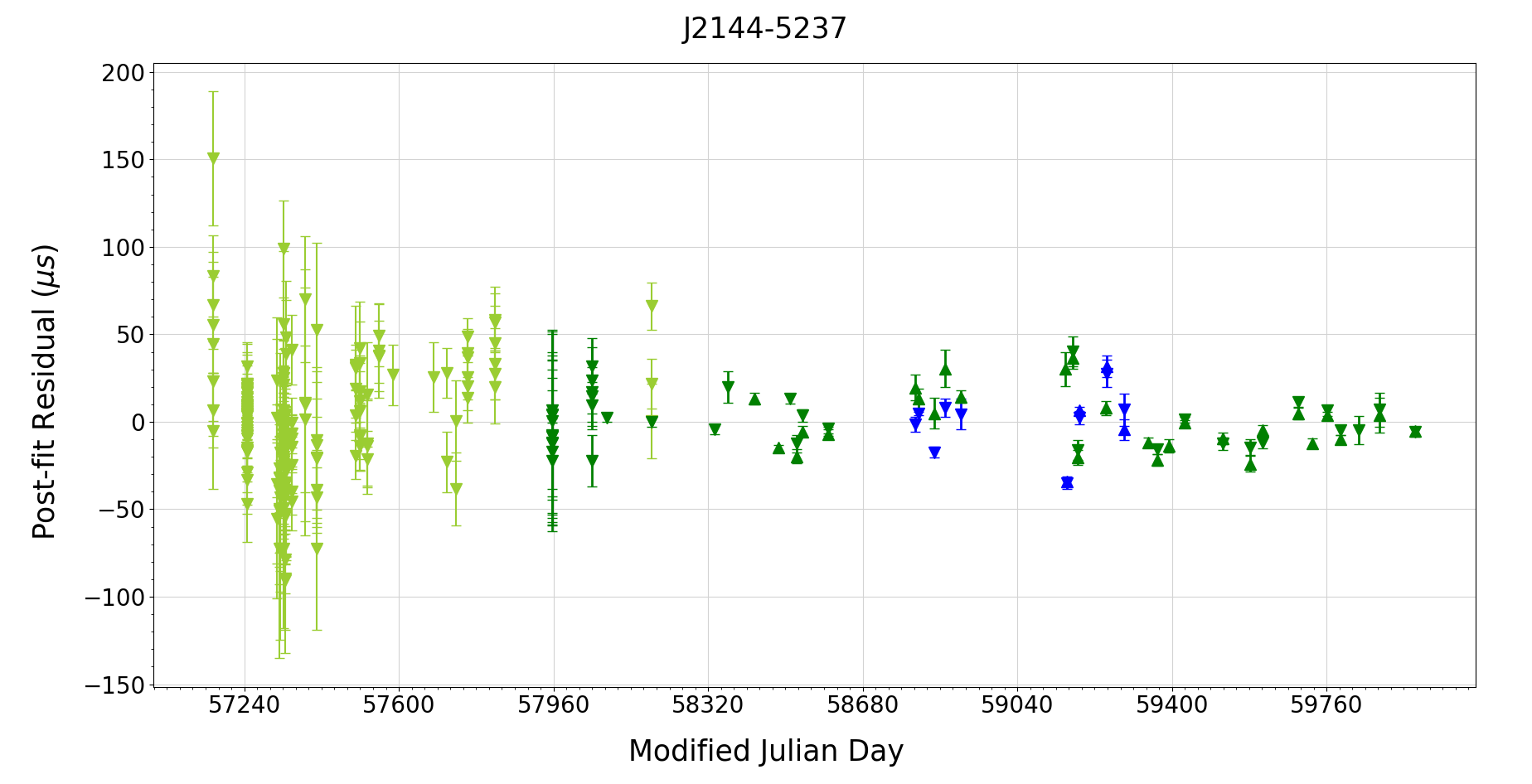}
        \vspace{-5mm}
    \caption{Post-fit timing residuals vs MJD for MSP J2144$-$5237. The rest of the plotting approach is the same as in Figure \ref{J1646_timing}. However, this MSP did not have GSB band-4 data available.}
    \label{J2144_timing}
\end{figure}

Individual GSB and GWB timing resulted in unreliable proper motion values with large error bars. We find proper motion for this MSP (9.8(5) $mas/yr$) for the first time using combined GSB+GWB timing. %In particular, the proper motion in RA and DEC is 5.4(3) and -8.2(5) $mas/yr$, respectively. 

{The NE2001 and YMW16 models (together with DM and position) gave distance estimates of 0.80 and 1.66 $kpc$ for this MSP, respectively. The resulting transverse velocities are 37 $km/s$ (for 0.80 $kpc$, NE2001) and 77 $km/s$ (for 1.66 $kpc$, YMW16), respectively. These numbers are comparable to the transverse velocities of MSPs J1120$–$3618 and J1828$–$0625, and suggest a symmetric supernova explosion for this MSP as well. The timing fit for this MSP resulted in a P1 value of 9.055(3) $\times$ 10$^{-21}$ s/s, from which the intrinsic spin period derivative is calculated to be $\sim$5.9 $\times$ 10$^{-21}$ s/s ($\sim$0.6 P1).}
%{Similar to J1120, estimate VT, compare with theoretical model prediction, estimate the intrinsic P1. Add these values in the ephimeris table.}

\begin{figure}[H]
\centering
        \includegraphics[width=0.49\linewidth,keepaspectratio]{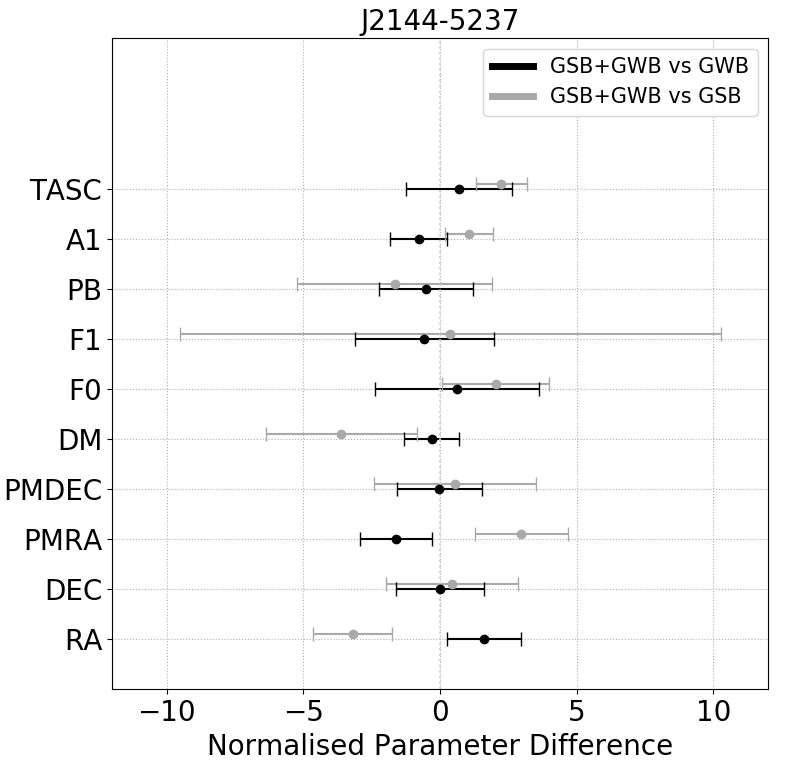}
        
    \caption{Figure showing a comparison of the fitted timing models for J2144$-$5237 in GSB+GWB timing to individual GWB and GSB timing. The plotting style is the same to Figure  \ref{J1120_model_plot}.}
    
    \label{J2144_model_plot}
\end{figure}

Figure \ref{J2144_model_plot} compares the model parameters obtained from GSB+GWB timing with those obtained from individual GWB and GSB timing. The model parameters obtained from GSB+GWB timing are on average 3.4 and 5.7 times more precise than parameters in individual GWB and GSB timing, respectively. Also, the model parameter values obtained from GSB+GWB timing are within $\pm$1 $\sigma$ of the model parameters obtained from individual GWB and GSB timing, where $\sigma$ is the error in model parameter obtained from individual timing of GWB or GSB data sets.

%%%%%%%%%%%%%%%%%%%%%%%%%%%%%%%%
\begin{table}[H]
    \centering
    \begin{adjustbox}{width=\linewidth,center}
    \begin{tabular}{|c|c|c|c|c|c|c|}
    \hline
    Parameter & Backend & Band & J1120$-$3618 & J1646$-$2142 & J1828$+$0625 & J2144$-$5237 \\
                   &        &                 &      &      &      &      \\
    \hline
    \hline
              & GWB & B3 & 33 & 43 & 32 & 51 \\ 
    $N_{ToA}$ &     & B4 & - & 23 & 8 & 13 \\
              & GSB & B3 & 32 & 192 & 273 & 185 \\
              &     & B4 & - & 27 & 8 & - \\
  
    \hline
     Median   & GWB & B3 & 6.74 & 4.09 & 4.51 & 3.28\\
    $\sigma_{ToA}$ &     & B4 & - & 5.97 & 4.18 & 4.32 \\
    ($\mu$s) & GSB & B3 & 19.52 & 16.56 & 22.19 & 20.55 \\
             &     & B4 &    - & 17.14 & 30.60 & - \\
      
    \hline

      Post-fit &   GWB  & B3+B4$^{\dagger}$  & 6.1 & 7.3 & 6.2 & 9.0 \\
    timing res- &  GSB &                         & 27.0 & 15.7 & 20.9 & 17.1 \\
    iduals ($\mu$s) &  GSB+GWB &              & 10.5 & 11.1 & 11.8 & 10.6 \\
                   &        &                 &      &      &      &      \\
 \hline
\hline
\end{tabular}

\end{adjustbox}

    \caption{The table lists the number of ToAs, median ToA uncertainties, and post-fit timing residuals for each MSP in both frequency bands and for both (and combination of) receiver backend data sets. B3 and B4 represent band-3 and band-4, respectively. The $\dagger$ on B4 in the third and last column and row is to indicate ``whenever available".}
\label{Median_ToA}
\end{table}

\begin{table}[H]
    \centering
    \begin{adjustbox}{width=\linewidth,center}
    \begin{tabular}{|c|c|c|c|c|}

 \hline
 Fitted parameters & J1120$-$3618 & J1646$-$2142 & J1828$+$0625 & J2144$-$5237 \\
                &            &            &            &            \\
 \hline
                 &            &            &            &            \\

 RA (hh:mm:ss.s) & 11:20:23.3553(4)& 16:46:18.6341(3) & 18:28:28.95501(5) & 21:44:35.6453(7) \\
 PMRA (mas/yr)   & 5.2(6)             &                  & -14.0(2)          & 5.4(3) \\
 DEC (dd:mm:ss.s) & -36:19:40.477(7)& -21:42:02.51(3) & +06:25:09.793(1) & -52:37:07.41(1) \\
 PMDEC (mas/yr)  & 2.8(7)              &                 & -3.0(2)          &  -8.2(5)  \\
 % PEPOCH &    56225.017213823751945   & 56596.524252000000001 & 57496.590724999999999 & 57328.500305999999998 \\
 % POSEPOCH & 56225.017213823751945 & 56596.524252000000001 & 57496.590724999999999 & 51544 \\
 F0 (s$^{-1}$)   & 179.95266943563(6)& 170.849405719173(5) & 275.667196397883(4) & 198.35548314683(2) \\
 F1 (s$^{-2}$)   & -3.03(9)e-17          &  -2.4248(3)e-16     & -3.5627(7)e-16      & -3.563(1)e-16 \\
 F2 (s$^{-3}$)   & 3.0(5)e-26        &                  &                      &               \\
 DM (pc$\,$cm$^{-3}$) & 45.128(3)    & 29.74000(9)         & 22.4165(1)          & 19.5500(2) \\
 BINARY          & ELL1               &    ISOLATED               &     DD               & ELL1          \\
 PB (days)       & 5.659945229(3)      &                 &  77.92496858(2)       & 10.580318287(5) \\
 A1 (lt-s)       & 4.304023(3)&                  & 34.888407(1)         & 6.361098(2)  \\
 T0 (MJD)        &                    &                  & 57546.579(8)         &               \\
 TASC (MJD)      & 56225.015639(2)    &                  &                      &  57497.7855778(7) \\
 OM (degrees)    &                    &                  & 230.35(3)            &                   \\
 ECC             &                    &                  & 9.795(8)e-05        &                    \\
 START-FINISH (MJD)   & 56220-59965  & 55869-59881      & 55869-59848          & 57168-59966        \\
 Reference MJD   & 57805              & 57924            & 58103                & 58539              \\
 NTOA            & 65                  & 285              & 321                  & 249                \\
 Number of Fit Parameters & 11                  & 6                & 13                   & 11                 \\
 Post-fit timing residuals ($\mu$s)  & 10.462             & 11.095           & 11.814               & 10.550             \\
 Units           & TCB                & TDB              & TDB                  & TDB                \\
 EPHEM           & DE200              & DE200            & DE405                & DE200              \\
 TIMEEPH         & IF99               & FB90             & FB90                 & FB90               \\
 \hline
Derived Parameters &                  &                  &                      &                    \\

\hline
 &                  &                  &                      &                    \\
P0 (ms)            & 5.557016759664(2) & 5.8531078629779(2) &  3.62756255755819(5) & 5.0414537785161(4) \\

Transverse velocity  (km/s) [YMW16] & 27 & & 68 & 77 \\
Measured P1  (s/s)         & 9.4(3)e-22          & 8.307(1)e-21       &  4.6882(9)e-21       & 9.055(3)e-21       \\
Intrinsic P1 (s/s) [YMW16] &  5.5e-22 & & 2.7e-21 & 5.9e-21 \\

%Characteristic age (Gyr) & 94.042 & 11.17             &  12.27              & 8.83               \\
Surface Magnetic field (10$^8$ Gauss)& 0.730        &  2.231             & 1.320               & 2.162              \\   
Median Companion Mass ($M_\odot$)   & 0.2161        &                   & 0.3180               & 0.2099             \\
Spin-down Luminosity (10$^{33}$ erg s$^{-1}$) & 0.2157   &   1.6364            & 3.8793              & 2.7914             \\
Total time span (years) & 10.253    & 10.987             & 10.894               & 7.662               \\
\hline
\end{tabular}
\end{adjustbox}
    \caption{The table lists the parameters that have been fitted or derived for a specific MSP along with the units of measurement (if applicable). The parameters and their uncertainty (if applicable) values are derived from band-3+band-4 (whenever available) and GSB+GWB timing fit. The parameter definitions are the same as in Table 2 and 3 of the tempo2 manual, which is available at \url{https://www.jb.man.ac.uk/~pulsar/Resources/tempo2_manual.pdf}.}
    \label{timing_solutions_full}
\end{table}

\subsection{Comparison with MSPs currently included in PTAs}

\begin{figure}[H]
\centering
        \includegraphics[width=1.0\linewidth,keepaspectratio]{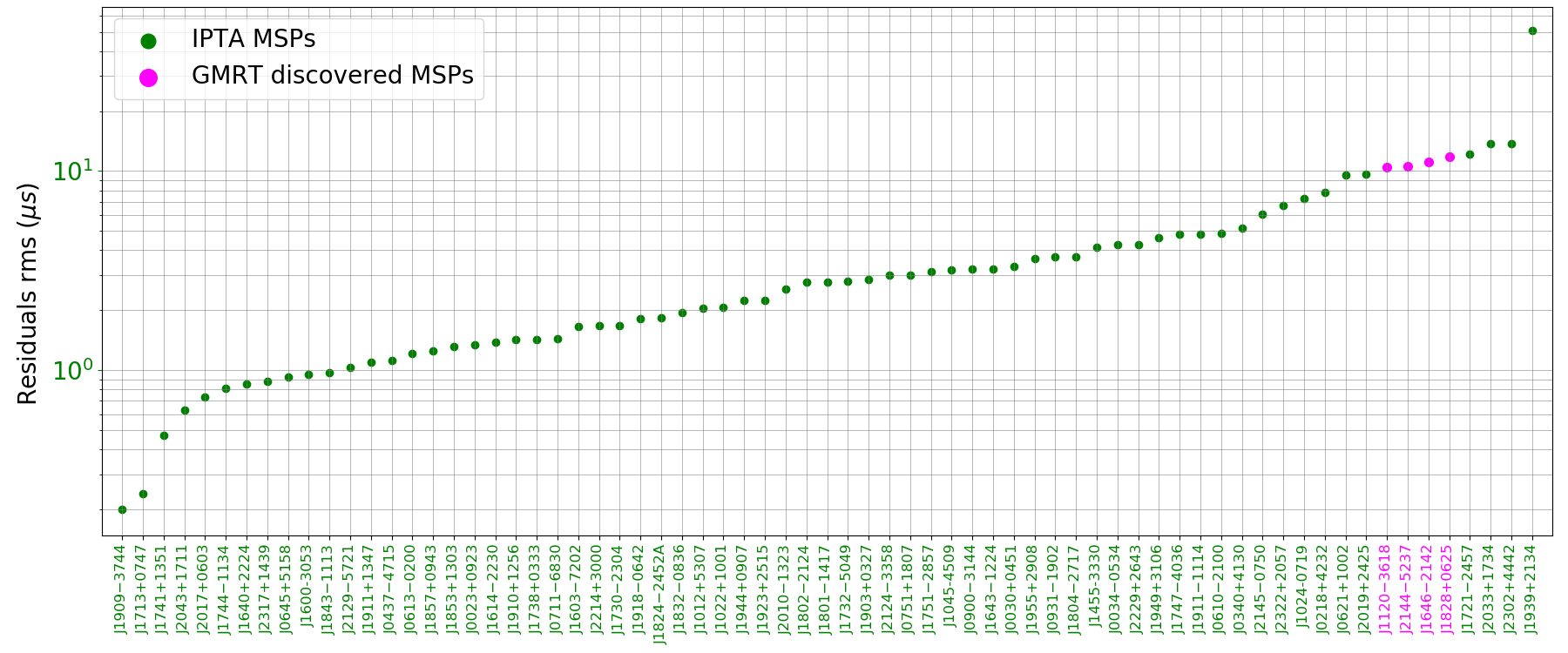}
        \includegraphics[width=0.8\linewidth,keepaspectratio]{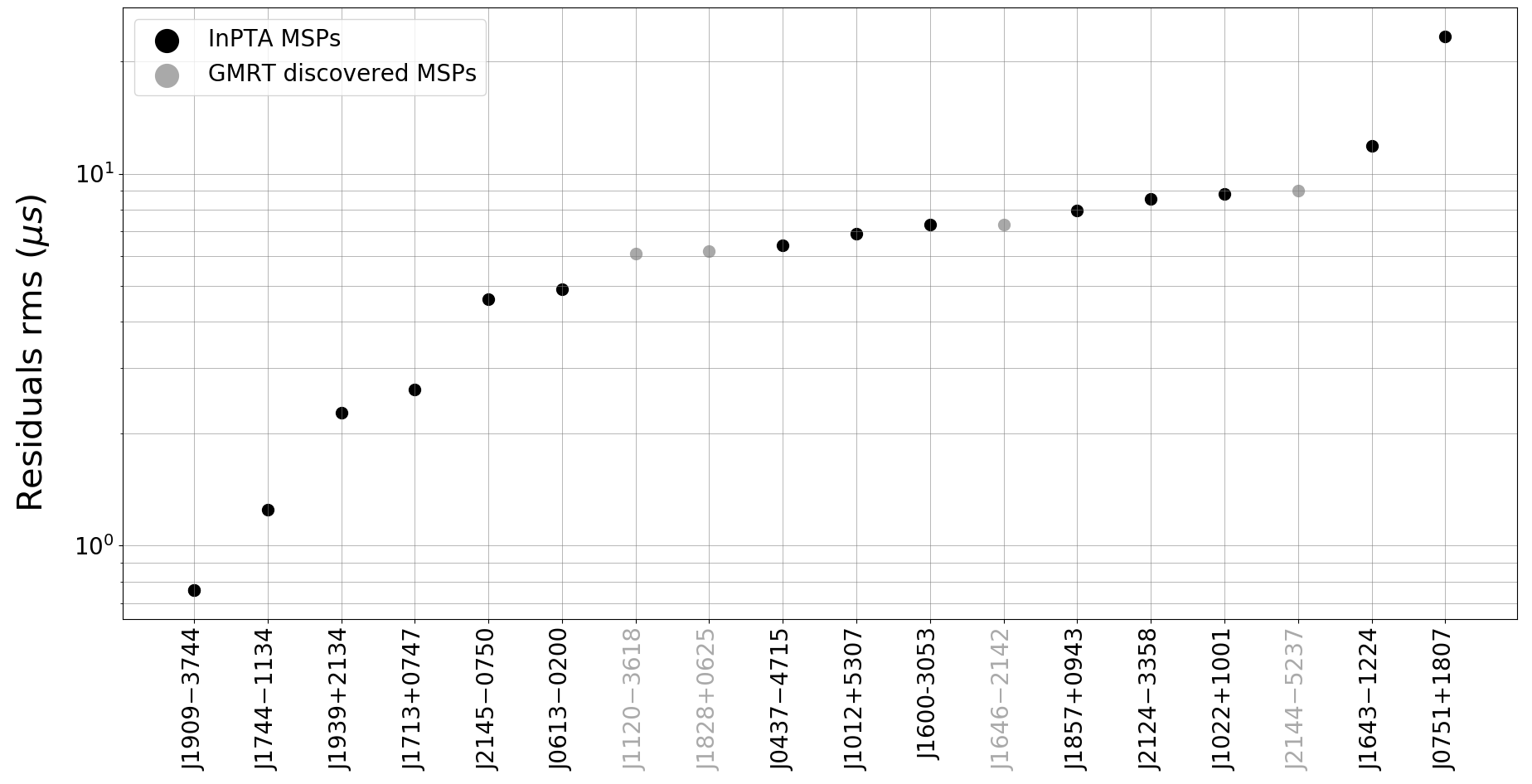}
       % \vspace{-5mm}
    \caption{{Top panel:} Figure showing the timing precision of 65 IPTA MSPs and four GMRT discovered MSPs. Here, we've used the timing precision for the IPTA MSPs (green color) listed in appendix A of \cite{2019MNRAS.490.4666P}. For the GMRT discovered MSPs, we plotted the timing precision (magenta color) obtained from GSB+GWB timing given in Table \ref{Median_ToA} of this work. On the x-axis, MSPs are sorted in decreasing order of timing precision. {Bottom panel:} Timing precision of 14 PTA MSPs monitored by InPTA (using uGMRT) and four GMRT discovered MSPs. Here, we've used the timing precision for the InPTA MSPs (black color) listed in Figure 5 of \cite{2022PASA...39...53T}. For the GMRT discovered MSPs, we plotted the timing precision (dark-gray color) obtained from only GWB timing given in Table \ref{Median_ToA} of this work. Again, MSPs are sorted in decreasing order of timing precision on the x-axis.}
    \label{Comparison_with_IPTA_MSPs}
\end{figure}

The International pulsar timing array (IPTA; \cite{2016MNRAS.458.1267V}, \cite{2019MNRAS.490.4666P}) combines timing data from three major PTAs: NANOGrav, PPTA, and EPTA, with the goal of improving timing precision for individual MSPs. IPTA second data release \cite{2019MNRAS.490.4666P} reported timing analyses of 65 MSPs, providing the model parameters and timing residuals obtained for each MSP. In Figure \ref{Comparison_with_IPTA_MSPs}, we plotted the RMS of timing residuals obtained for the 65 IPTA MSPs and 4 GMRT MSPs in GSB+GWB timing. 

The 4 GMRT discovered MSPs fall inside timing precision range for the IPTA MSPs. %specifically, their timing precision ranks ranged from 62 to 65 out of a total of 69 (65 IPTA $+$ 4 GMRT) MSPs. 
We note that the RMS timing residuals obtained from individual GWB timing are (on average) 1.5 times less than the RMS timing residuals obtained from GSB+GWB timing. Therefore, considering solely the GWB timing precision makes them even more promising candidates for PTA experiments. {Moreover, \cite{2022PASA...39...53T}(InPTA) reported the RMS timing residual of 14 well-timed PTA MSPs (observed with uGMRT) from GWB timing observations for a span of $\sim$ four years. The timing precision for these bright PTA MSPs at the uGMRT ranges between 0.76$-$23.4 $\mu$s (Figure \ref{Comparison_with_IPTA_MSPs}; bottom panel). This clearly illustrates that the timing precision of the four GMRT-discovered MSPs (based on GWB observations) is well within the range of PTA MSPs over a similarly sized timing span.}  
 
\section{Summary}

In this work, we compared the timing results of the 4 GMRT-discovered MSPs for GWB (spanning 3.1$–$5.5 years), GSB (spanning 1.8$–$6.1 years), and GSB+GWB (covering 7.7$–$11.0 years) combined observations. The RMS of timing residuals obtained from GSB+GWB timing is on average $\sim$1.6 times higher than individual GWB timing but $\sim$1.8 times (on average) lower than the residual's RMS obtained from GSB timing. The fitted model parameters in GSB+GWB timing are on average 5 and 4 times more precise than those in individual GWB and GSB timing, respectively. Model parameters from GSB+GWB timing are consistent with individual GWB and GSB timing within $\pm 1$$\sigma$, where $\sigma$ is the error in model parameter obtained from individual timing of GWB or GSB observations.

We presented DM variations for the 4 MSPs derived from GWB observations. We find that the line of sight for J1120$-$3618 crosses an electron-rich medium. We see an $\sim$9$\,\sigma_{DM}$ increase in the DM value for this MSP during the GWB observations period of 4.2 years, where $\sigma_{DM}$ is the median DM uncertainty. For the remaining three MSPs, the change in DM values is contained within $\pm3$ $\sigma_{DM}$.

We detected proper motion for the first time for J1120$-$3618 (6.0(6) $mas/yr$) and J2144$-$5237 (9.8(5) $mas/yr$). For J1828$-$0625, we detect a proper motion of 14.4(2) $mas/yr$, which is at least five times more precise than that reported in \cite{2022ApJ...933..159B}. %The proper motion values for these three MSPs fall inside the normal proper motion range for the known MSPs.
{The transverse velocity for the three MSPs, calculated from these proper motion values and the distance estimates derived from NE2001 and YMW16, falls within the range expected for the MSPs (e.g., \cite{2005MNRAS.360..974H}).}

Finally, we compared the timing precision achieved for the four GMRT MSPs to that of 65 IPTA MSPs reported in the second IPTA data release \citep{2019MNRAS.490.4666P} and with 14 MSPs reported in first data release of InPTA \citep{2022PASA...39...53T}. The GMRT MSPs fall within the timing precision range of the IPTA MSPs and InPTA observed MSPs, making them potential candidates for PTA experiments. The decade-long timing data for the four GMRT MSPs utilised in this study may be useful in the ongoing global effort to aid to the detection significance of the common correlated signal recently detected by multiple PTAs.

\label{sec:Section-5}

\section{Acknowledgments}
We gratefully acknowledge the Department of Atomic Energy, Government of India, for its assistance under project no. 12-R$\&$D-TFR-5.02-0700. The GMRT is run by the National Centre for Radio Astrophysics - Tata Institute of Fundamental Research.

\end{document}